\documentclass[a4paper,twoside]{article}

\usepackage{epsfig}
\usepackage{subcaption}
\usepackage{calc}
\usepackage{amssymb}
\usepackage{amstext}
\usepackage{amsmath}
\usepackage{amsthm}
\usepackage{multicol}
\usepackage{pslatex}
\usepackage{apalike}
\usepackage[bottom]{footmisc}

\usepackage[]{graphicx}

\usepackage[ruled,vlined, linesnumbered]{algorithm2e}

\usepackage{xcolor}
\usepackage{subcaption}

\usepackage{SCITEPRESS}     % Please add other packages that you may need BEFORE the SCITEPRESS.sty package.

\begin{document}

\title{PhilaeX: Explaining the Failure and Success of AI Models in Malware Detection}

\author{\authorname{Zhi Lu and Vrizlynn L. L. Thing}
\affiliation{Cyber Security Strategic Technology Centre, ST Engineering, Singapore}
\email{{lu.zhi@stengg.com}, {vriz@ieee.org}}
}


\abstract{The explanation to an AI model’s prediction used to support decision making in cyber security, is of critical importance. 
It is especially so when the model’s incorrect prediction can lead to severe damages or even losses to lives and critical assets. 
However, most existing AI models lack the ability to provide explanations on their prediction results, despite their strong performance in most scenarios.
In this work, we propose a novel explainable AI method, called PhilaeX, that provides the heuristic means to identify the optimized subset of features to form the complete explanations of AI models’ predictions. 
It identifies the features that lead to the model’s borderline prediction, and those with positive individual contributions are extracted. 
The feature attributions are then quantified through the optimization of a Ridge regression model. 
We verify the explanation fidelity through two experiments. First, we assess our method’s capability in correctly identifying the activated features in the adversarial samples of Android malwares, through the features attribution values from PhilaeX. Second, the deduction and augmentation tests, are used to assess the fidelity of the explanations. 
The results show that PhilaeX is able to explain different types of classifiers correctly, with higher fidelity explanations, compared to the state-of-the-arts methods such as LIME and SHAP.}

\onecolumn \maketitle \normalsize \setcounter{footnote}{0} \vfill

\section{\uppercase{Introduction}}
\label{sec:introduction}

Explaining the prediction of an AI model is critical for the AI-based solution to modern cyber threats that have the properties of large volume and highly complexity by the AI technology.
The threat detection solutions based on the learnable AI technologies, which are so called shallow machine learning and recently emerging deep learning methods, have demonstrated astonishing performance today.
However, the high detection performance is insufficient in establishing the trust from the users, since most models predict the label of the suspicious sample, e.g., a malware or a face image may be subjected to manipulation for deception or obfuscation, through a complicated computation process that people cannot understand.
This confidence crisis may become more severe when the AI model makes an erroneous prediction that causes damage or loss to the user's properties, assets or even safety.
Therefore, the research on explainable AI that quantitatively explains the AI model's successful or failed prediction for a particular input sample through the attribution of each data feature's contribution to the model's prediction is highly desired~\cite{dovsilovic2018_xai_survey}.

Malware detection research has made progress over the years. 
Demontis et. al.~\cite{demontis2017yes} improved the standard SVM on Android malware detection that further reduces the chance of evasion by certain types of malware samples, through the optmized selection method on the model's parameters.
Zhang et. al.~\cite{zhang2019scalable} proposed a malware detector using online learning technique that is capable of adapting to the rapid evolving malware. Specifically, they combined the n-gram analysis and the online classifier techniques in the detection.
The application of the deep learning methods in cyber security threats detection recently, such as CNNs~\cite{amerini2019deepfake}, RNNs~\cite{guera2018deepfake}, LSTM~\cite{xiao2019android} or Transformers~\cite{devlin2018bert}, is a breakthrough in the detection rate (i.e, true positive rate).
The deep learning methods also save the hand-crafted and time-consuming efforts on the selection or transformation of the samples' features through the automatic end-to-end learning, which performance was highly based on the experience and the domain knowledge of the developers~\cite{mclaughlin2017deep}~\cite{yan2018lstm}~\cite{android_malware_LSTM_2019} previously.
However, it is nearly impossible for humans to understand how the deep learning models predict the class of the samples by the non-linear computation process and millions parameters among layers.
The research effort on AI models' explanation is seldom considered in the development of the machine learning algorithm.

Clearly, the AI model explanation is the positive direction to enhancing the users’ trust on the AI  model’s  output, otherwise generated from a seemingly black-box mechanism.
Such explanation is achieved through the quantification on the “contribution” of each feature to the model’s prediction.
The popular model-agnostic explainable AI methods that can explain any AI model's predictions, regardless of the model's type (such as SVM, CNNs or LSTM), may not be working well for cyber security problems.
LIME~\cite{lime} builds a surrogate linear model of the original model to be explained, where the contribution of each feature is computed through the optimization~\cite{efron2004least}.
The authors assumed the linear model can be understood by humans because of its simplicity and the data used to train the linear model is manipulated by the local perturbation of the features values in the input data sample.
The fidelity of the linear model based explanation may be deteriorated by the high dimensionality of the data that is common in cyber security.
Integrated Gradients (IGs)~\cite{integrated_gradients} attributes the features as the model explanation through the integration of the gradients on the model's predictions with respect to the input data with different features values.
These feature values are varied from the ``baseline'' through a linear path, in which the baseline refers to the zero-value feature vector or no signal sample.
The Integrated Gradients method works well for the AI models with gradients, such as deep learning models. 
However, it cannot be used for certain widely used models without gradients, such as Random Forests~\cite{apruzzese2020hardening}.
In addition, the baseline is unclear in certain fields, such as genomics domain~\cite{jha2020enhanced}.
Therefore, the explainable AI method for the models used in the cyber security field, such as malware detection, is still desired.

In this article, we proposed a novel model-agnostic explainable AI methodology, called \textit{PhilaeX}, that is capable of quantitatively measuring the features' ``contribution'' in a suspicious app sample, when its class (i.e., benign or malware) is predicted by a given AI model, regardless of the model's type.
Specifically, the model explanation starts from \textbf{core features} selection for a given suspicious sample, by which only the features in the sample lead the model's prediction towards to the border line of the two classes (i.e., around $50\%$ probability of the prediction confidence by the model) are selected.
Then, in addition to these core features, PhilaeX identifies a set of features from the original data sample, in which each feature is able to make the significant contribution for the model's prediction towards the predicted class on the original input sample.
This step is to identify the \textbf{features with positive individual contributions} to the model's predictions, without considering the contributions from the cooperation among features.
Finally, the \textbf{feature attribution} is obtained by considering both the positive individual contributions and the joint contribution when all these features are used.
The quantitative measure on each feature's attribution is computed by optimizing a Ridge regression, because of its simplicity in optimization and the nature of the optimization considers the highly correlated features.
The \textbf{main advantages} of the proposed explainable AI method include:
(1) The identification of the core features provides a fingerprint to further identify the candidate features with positive contributions to the model's prediction, in an efficient and accurate manner, when compared to the random perturbation of the sample's values in feature space, such as LIME;
(2) The features attribution based on the core features and those with positively individual contributions considers both the individual and joint contributions by the features;
and (3) The optimization by Ridge regression to quantify the features attribution is efficient and effective.
The results from the quantitative assessment to the proposed explainable AI method show the high fidelity of explanation by PhilaeX, regardless of the SVM~\cite{drebin_dataset}~\cite{li2015detecting} and BERT~\cite{devlin2018bert} classifiers, on malware detection tasks.
The first experiment aims to identify the ``activated features'' in adversarial samples of Android malware.
This is to help the cyber security practitioners to analyze how the AI model was evaded by the adversarial samples, and enhance the model's security accordingly.
The results demonstrate that the activated features have the higher chance to be attributed with high values by PhilaeX, compared to the state-of-the-arts methods, such as LIME, SHAP~\cite{NIPS2017_7062} and MPT Explainer~\cite{mpt_explainer_2021}.
The second experiment that test the explanation fidelity when PhilaeX is used to explain the SVM and Random Forest classifiers on the PDF malware dataset~\cite{smutz2012malicious}, where the results verifies that the high fidelity can be obtained by a small number of the features with the top attribution values by PhilaeX.

The rest of the paper is organised as follows:
we present literature review on the state-of-the-arts in explainable AI in cyber security in Section~\ref{sec_lit_review}. 
The proposed methodology, PhilaeX, is introduced in details in Section~\ref{sec_method_philaex}.
We assess the fidelity of the proposed method by two quantitative experiments in Section~\ref{sec_experiment}. 
Finally, the conclusion of this methodology is discussed in Section~\ref{sec_conclusion}.

% +------------------------------+
% |      Literature Review       |
% +------------------------------+
\section{Literature Review}
\label{sec_lit_review}

The main aim of explainable AI is to provide a human-understandable explanation on how the AI model predicts the class label of the given sample.
One of the major research in explainable AI focus on the model's interpretability~\cite{dovsilovic2018_xai_survey}, where the model's prediction can be explained by its own prediction process, such as decision trees~\cite{kaminski2018framework}.
However, as the development of the machine learning and deep learning methods advances, the model becomes increasingly complicated such that the computation is not visible for the users, and it is difficult to achieve the model's interpretability~\cite{molnar2019}.

The \textit{post-hoc} explainable AI methods that obtain the model's explanation by analyzing the model's input and output in a qualitative or quantitative way, therefore, attracts the major research interests.
The early research on post-hoc explaination method were focusing on the \textit{model-specific} explainable AI methods, where it is only able to explain the targeted type of AI models.
Zeiler~\textit{et. al.}~\cite{zeiler2014visualizing} proposed a qualitative explanation method through the visualization and observation on the neurons in a convolutional neural networks (CNNs) that shows how each neuron responds to different data instances.
In~\cite{xu2015show}, Xu~\textit{et. al.} developed a caption generator model to summarize the content of an image in one sentence, where the attention mechanism in the deep neural networks highlights the sensitive part of the image and its corresponding words in the caption.
DREBIN~\cite{drebin_dataset} provided a limited explanation of the Android malware detector's prediction based on SVM classifier. However, their explanation method cannot be extended to other AI models, since the quantification of features attribution comes from the weights of the SVM models.
Thus, the model-specific explainable AI methods lack the ability to extend to new types of AI models, because of its inherent nature.

As the machine learning techniques develop rapidly, explainable AI methods that can explain different types of AI models is highly desired. This property is also referred to as \textit{model-agnostic}.
Samek et. al.~\cite{samek2017explainable} firstly proposed the explanation methods using \textit{layer-wise relevance propagation} (LRP) to analyze the sensitivity between the deep learning models' prediction w.r.t. the input sample in the features space.
Their work forms a foundation in model-agnostic explainable AI methods, where the model explanation was obtained by the ``observation'' on the relations between the input and model's output. 

As the model structure becomes too complicated to be accessed by humans, the directly observation on the model's input and output also become a time-consuming and inaccurate way to obtain the explanation.
Therefore, the alternative way to obtain the model explanation is to explain the surrogate model, which simulates the behavior of the original model to be explained, and is usually simple enough for human understanding.
LIME~\cite{lime} is proposed to explain any type of classifiers by learning a linear surrogate model to mimic the target model's behavior.
The data to train such linear model are generated through perturbation of the original input data sample around the model's predictions (i.e., local perturbation).
However, the linearity of the surrogate model and the random perturbation strategy in the local field limits the explanation capability of LIME, especially when it explains complicated models, such as CNNs.
Our PhilaeX provides a high fidelity explanation for complicated models through a multi-stage selection strategy for high contribution features. This solves the limitation of the local random perturbation in the sample's feature space, such as non-stable explanation.
Wu~\textit{et. al.}~\cite{wu2018beyond} used decision tree, which is a self-explained model, as the surrogate model in the explanation of the deep learning models
Recently, LEMNA~\cite{guo2018lemna} was proposed to explain the AI models that are specifically designed for cyber security problems.
LEMNA uses the \textit{fused lasso}~\cite{tibshirani2005sparsity} algorithm and \textit{mixture regression model}~\cite{khalili2007variable} to force the explanation to consider the dependencies among features, which solves the issues of linear approximation that considers nothing about the dependencies among features in LIME.

% +------------------------------+
% |         Methodology          |
% +------------------------------+
\section{PhilaeX: Explaining Model's Predictions}
\label{sec_method_philaex}
In this section, we firstly formulate the model's explanation problem as the feature attribution process in mathematics.
The algorithms to identify the \textbf{core features} and the features with positive individual contributions are introduced.
Finally, we present the optimization process to obtain the features attribution by considering both the features' individual contributions and their joint contributions towards the model's prediction on the input sample.

\subsection{Problem Statement}
\label{sec_sub_problem_statement}
Given a classifier $f(\mathbf{x}) \rightarrow [0,1]^{|C|}$ to be explained, and its predictions of the probabilities of $|C|$ class labels for the input data sample $\mathbf{x}=(x_1, x_2, ..., x_m) \in R^m$ (in the features space), the data sample $\mathbf{x} \in R^m$ consists of $m$ features. 
For example, the suspicious Android app can be represented in the features space by the TF-IDF~\cite{rajaraman2011data} values of its permissions~\cite{drebin_dataset}.
The aim of the model explanation is to find the optimized features attribution vector $\mathbf{A} = ( a_1, a_2, ..., a_m ) \in R^m$ that quantitatively measure how the model to be explained $f(\mathbf{x})$ makes the prediction of the input sample's class label according to each features contributions.
That is, the optimization can be formally represented by:

\begin{equation}
    \mathbf{A} = argmin_{\mathbf{x^*}}(g(h(\mathbf{x}), w) - f(\mathbf{x}))
\end{equation}

where $g(\cdot) \in \mathcal{G}$ is the surrogate model to the original classifier $f(\cdot)$, which aims to mimic the predictions as $f(\mathbf{x})$ for the same sample $\mathbf{x}$, and the weights $w$ measures the joint contributions by the features in this surrogate model.
The selection function $h(\mathbf{x})$ returns the optimized features set $\mathbf{x^*}$ that make the significant contributions to the model's predictions, given sample $\mathbf{x}$.
The attribution vector $\mathbf{A}$ is obtained only if the minimized difference between the surrogate model $g(\mathbf{x})$'s prediction and that of the original model $f(\mathbf{x})$ to be explained is obtained.
Therefore, the choice of the surrogate model and the features $\mathbf{x^*}$ that their attributions are computed is critical to the explanation fidelity on the model's prediction behavior on the sample $\mathbf{x}$.

In the remaining part of this section, we will introduce our proposed model explainer, PhilaeX, that starts the features attribution vector construction for the input sample $\mathbf{x}=(x_1, x_2, ..., x_m) \in R^m$ from an empty vector (i.e., Null).
The whole construction process consists of two major stages:
(1) The features selection strategy, i.e., $h(\mathbf{x}) \in R^n, n \leq m$, that picks up the features with the significant contributions towards the model's prediction on $\mathbf{x}$ is selected;
(2) The quantification of the contribution for the selected features through a Ridge regression that is the surrogate model to the original model $f(\mathbf{x})$.

\subsection{Core Features}
\label{sec_sub_core_features}

The perturbation of the features values to obtain the synthetic input data samples $\mathbf{X^{'}}$ in the training of the surrogate model $g(\mathbf{X^{'}})$ may not work well in the cyber security field.
In LIME~\cite{lime}, the response of the model to the changes of the input variables is obtained by random perturbation of the input sample's feature values in a small range.
This can allow a fast preparation of a large amount of synthetic data to train the surrogate linear model and help the explainer to attribute the model's behavior accordingly.
However, it can also lead to the shortage of stable explanation that the features attribution values may vary a little among different times of explanation, given the same input sample.
In addition, the perturbation strategy on features magnitude to generate the synthetic data to train the surrogate model may not work well in the cyber security field.
For example, the normal way to camouflage the malware to evade the AI detector is to ``add'' a certain types of permission in the app, where the small perturbation of the features values is not impossible.

In PhilaeX, the features selection function $h(\mathbf{x}) \in R^n, n \leq m$ is to pick up the subset of the features from the input sample $\mathbf{x}$ that is optimized to describe (i.e., explain) the model's prediction behavior.
Specifically, there are two steps to obtain the candidate features for attribution, which are \textit{core features} and \textit{features with positive individual contributions}, respectively.
The first step is to identify a set of \textbf{core features} $\mathbf{x_c} = \{x_i \in \mathbf{x}\}$ from the original sample $\mathbf{x}$, which are the \textit{base} of the sample $\mathbf{x}$ that leads the model $f(\mathbf{x_c}) :\to 0.5$ (i.e., the boarder line of the prediction).
We assume that the model $f(\mathbf{x_c})$ make a ``hesitated'' decision for the sample with such core features only, where the model has around 50\% confidence on its prediction of the sample's class, and the actual prediction on the original input sample $f(\mathbf{x})$ is made by the joint contribution from both the core features and part of the remaining features.

\begin{algorithm}
\SetAlgoLined
\SetKwInOut{Input}{Input}
\SetKwInOut{Output}{Output}
\Input{
Input sample: $\mathbf{x}$ and model to explain: $f(\cdot)$
\newline
MAX\_LEN\_CORE\_FEATURES - maximum number of core features
}
\Output{
    Core features: $\mathbf{x}_s$
\newline
}

$min\_prediction\_score\_gap = 1$

$\mathbf{x}_s = Null$

\While{$\|\mathbf{x}_s\| \leq MAX\_LEN\_CORE\_FEATURES$}{

    Pick up $x_i \in \mathbf{x}$

    \uIf{$\|f(\mathbf{x}_s + x_i) - 0.5\| < min\_prediction\_score\_gap$}{
        $x_i :\to \mathbf{x}_s$
    }
}

\Return The selected core features $\mathbf{x_s}$

\caption{Core Features Selection}
\label{alg:the_alg_xstar}
\end{algorithm}

In order to obtain the core features for the given sample $\mathbf{x}$, we start from an empty feature vector that contain no feature.
The following steps are to find out the candidate core features in a recursive way, where the target is to find the subset of features that leads the model $f(\mathbf{x}_c) :\to 0.5$ as close as possible (i.e., the local minimum of the $abs(f(\mathbf{x}_c) - 0.5)$.
The detailed algorithm about core features identification are in Algorithm~\ref{alg:the_alg_xstar}.

\subsection{Features Individual Contributions}
\label{sec_sub_features_individual_contributions}
Once the core features $\mathbf{x}_c$ is obtained, we are looking for the features that can increase the prediction confidence of the model toward the prediction score on the original sample $\mathbf{x}$.
Formally, we define the acquisition of such features with \textit{positive individual contributions}, i.e., $\mathbf{x}_p$, as:

\begin{equation}
    argmin_{\mathbf{x}_p \subset \mathbf{x} \backslash \mathbf{x}_c}{\left(f(\mathbf{x}_c + \mathbf{x}_p) - f(\mathbf{x})\right)}
\end{equation}

where the symbol ``$+$'' means the concatenation of two features vectors, i.e., $\mathbf{x}_c$ and $\mathbf{x}_p$.
The candidate features set is initialized as $\mathbf{x}_p = \phi$.
For every feature $x_i \in \mathbf{x} \backslash \mathbf{x}_c$ that is added into $\mathbf{x}_p$, the model's prediction on $f(\mathbf{x}_c + \mathbf{x}_p :\to f(\mathbf{x})$.

The aim is to identify the features in the input sample $\mathbf{x}$ to enhance the confidence of the model significantly when it outputs the prediction of the input sample.
Accordingly, those features that lead the model to the opposite of the prediction on the sample $\mathbf{x}$ will be ignored.

\subsection{Quantify Joint Contribution by Features}
\label{sec_sub_quantify_Joint_contribtuion_by_features}
The features we picked up from the previous steps, i.e., the core features $\mathbf{x}_c$ and the features with positively individual contributions $\mathbf{x}_p$, form the set of the candidate features, where features attribution by PhilaeX will be applied.
There are two reasons that we only attribute the subset of features in the input sample $\mathbf{x}$:
(1) The features attribution on such features $\mathbf{x}_s = \mathbf{x}_c + \mathbf{x}_p$ allows the explainer to reveal the major reason that the model made the prediction on the original sample $\mathbf{x}$. As the discussion in Section~\ref{sec_sub_features_individual_contributions}, it is not always true for all features in the sample $\mathbf{x}$ that make positive significant contributions towards the model's prediction of the class label on the sample.
(2) The explanation on such subset of features will be more efficient that that on the all the features of the input sample $\mathbf{x}$.

The joint contributions made by the cooperation among these features are the necessary to form the complete quantitative explanation (i.e., features attribution) for the model $f(\mathbf{x})$, which have not yet been considered by the previous two steps in Section~\ref{sec_sub_core_features} and Section~\ref{sec_sub_features_individual_contributions}.
In this step, we quantify each feature's contribution to the model's prediction by training a Ridge regression model $g(\cdot)$ as the surrogate model to the original model $f(\cdot)$, where the weights of each feature in the regression model are considered as the features attribution.
The reason we use the Ridge regression as the surrogate model is for its simplicity, efficiency and its nature for estimating the coefficients (i.e., weights) where independent variables are highly correlated~\cite{hilt1977ridge}.

Specifically, the weights $\mathbf{w} \in R^{\|\mathbf{x}_s\|}$ in Ridge regression can be estimated by the optimization of the following equation:

\begin{equation}
    argmin_{\mathbf{w}}{\left(||y - \mathbf{Xw}||^2_2 + \alpha * ||\mathbf{w}||^2_2\right)}
\end{equation}

where the L2 regularization applies to reduce sensitivity to single feature and accordingly decrease the possibility of overfitting in the model training.

Finally the features attribution vector is defined as $\mathbf{A} = \mathbf{w}$ that considers both the individual contribution from each features and the joint contributions from the cooperation among these features $\mathbf{x}_s$.

% +------------------------------+
% |         Experiments          |
% +------------------------------+
\section{Experiments}
\label{sec_experiment}
In this section, we assess the explanation capability of PhilaeX through two quantitative experiments.
The proposed explainer will be used to explain the prediction behaviors of three classical classifiers, including SVM, Random Forest and BERT, which include the AI models in both the shallow (classical) machine learning and deep learning fields.
There are two datasets are used in our experiments.
The datasets are DREBIN~\cite{drebin_dataset} dataset for Android malware detection task and the PDF malware dataset~\cite{smutz2012malicious} for PDF malware detection.
The explanation performance will be evaluated quantitatively in terms of the explanation fidelity in two tasks, which are the activated features identification for adversarial samples of Android malware and the deduction/augmentation tests for PDF malware samples.

\subsection{Dataset}
\label{sec_sub_exp_dataset}
We use two datasets in the evaluation on the explanation fidelity by PhilaeX.
The first dataset, DREBIN~\cite{drebin_dataset}, was used to test a lightweight Android malware detector, where the features of the suspicious Android apps were extracted from the application's manifest file \textit{AndroidManifest.xml} and  disassembled \textit{dex code} from the bytecode by the \textit{static analysis technique}.
The features that DREBIN extracted fall into 8 categories, like requested permissions, restricted API calls and network addresses, etc.
In the DREBIN dataset, there are 5,560 Android malware apps and 123,453 non-malware apps in total.
However, in our experiments, we randomly selected 5,555 malware samples and 5,555 non-malware apps, in order to build a balanced dataset for the model's training.
Further, the dis-joint training set and testing set used in the evaluation are built through a random split of these 11,110 samples, which generates a training set of 7,442 samples and a testing set of 3,668 samples.
For each sample, the text features data in a sample will be converted into the features vector in the form of floating numbers.
Specifically, all the features in the training dataset will be encoded by the \textit{tf-idf} algorithm~\cite{rajaraman2011data}, that measures the importance of each feature in the dataset.
The dimension of the features vector is 43,157, which is high dimension.

The second dataset used in the experiments is the PDF malware dataset~\cite{smutz2012malicious} that has 4,999 malicious samples and 5,000 benign samples.
We use the 135 features suggested by~\cite{guo2018lemna}, where the features have been encoded into binary (i.e., 0 or 1) values.

\subsection{AI Models to be Explained}
\label{sec_sub_ai_models_tobe_explained}
We test the explanation capability of PhilaeX for different AI models that cover the shallow (classical) machine learning and the recent emerging deep learning models.
First, we trained a SVM~\cite{drebin_dataset}~\cite{zhao2011antimaldroid}~\cite{li2015detecting} model, which is a classical shallow machine learning model and has been widely used as the classifier for binary classification tasks before the deep learning methods dominate this field.
For a given sample in the feature space, SVM maps the relatively low dimension data into a high-dimension space such that the separation between two classes becomes more apparent, and thus is able to predict the sample's class more accurately.

Specifically, we trained a SVM model with the Radial basis function (RBF) kernel~\cite{vert2004primer}, where the parameter that defines the inverse degree of the influence by a single training sample $\gamma$ is set to $1.0$.
We trained two SVM classifiers for the Android malware detection task on the DREBIN dataset and the PDF malware detection task on the PDF malware dataset.
In the remaining part of this section, we will use PhilaeX to explain the prediction behavior of these two classifiers (i.e., AI models).

In addition, we also trained a deep learning model for the Android malware detection task.
BERT~\cite{devlin2018bert}, the transformer-based classifier that was proposed by Google for natural language processing (NLP) tasks in 2018, is used to classify the Android malware in the DREBIN dataset.
We use the BERT implementation from HuggingFace Transformers library~\cite{Wolf2019HuggingFacesTS} that is not sensitive to the letters case and the default parameters, such as the maximum length of text (128) and the learning rate (4e-5).
There are 8 samples used in a single batch and 5 epochs were running in the training process of the BERT model.
We trained a surrogate SVM model to the BERT Android malware detector in the model explanation, in order to avoid the complicated word embedding mechanism that converts the text tokens to numerical representations.
Such surrogate SVM has highly similar prediction behavior as the BERT, given the sample input sample, where the TPR = 0.9984 and FPR = 0.0029.

Both the trained SVM and BERT models used in the Android malware detection tasks present good performance.
The true positive rate (TPR) for both classifiers are around 0.96 with a 0.04 false positive rate (FPR).
In addition, we also trained a separate SVM classifier and Random Forest classifier for the PDF malware detection task, which uses the default parameters.

\subsection{Explaining Evasion Attack by Adversarial Samples}
\label{sec_sub_quanti_assess_explain_adv_samples}
We firstly evaluate the explanation capability of PhilaeX on how the adversarial samples of Android malware evade the trained malware detector (that was with high TPR and low FPR on DREBIN dataset) in quantitative way.
In the evasion attack, we assume the attacker has full knowledge of the features space and access to the model's prediction score.
That is, the attacker is able to manipulate the data sample, which class is to be predicted by the SVM or BERT classifier, such as adding the features in the sample.

In this experiment, we only add (i.e., activate) the ``permission'' features to the existing sample's feature vector of Android malware, because such addition operation will not change the functionality of the original malware~\cite{liu2019adversarial}.
One adversarial sample is generated by Genetic Algorithm that is extended from~\cite{liu2019adversarial} and the optimised set of ``permission'' features is selected to help the original sample bypasses the malware detection by the classifier.
Specifically, in the Genetic Algorithm, the fitness value is defined by the model's prediction score towards the non-malware class of the candidate adversarial sample.
The convergence of the algorithm is fulfilled if (1) the Genetic Algorithm that has been running for 500 loops that has a high possibility to make the evasion attack by the adversarial samples successful; (2) the prediction score towards the non-malware class stay the same at a high level for at least 10 times; or (1) the fitness value is larger than 0.99 which implies the model has extremely high confidence on its incorrect prediction for the adversarial sample.
In total, there are 200 malware samples from the testing set randomly selected as the seeds to generate the adversarial samples.
The adversarial samples dataset used in the explanation for SVM has 499 samples.
In the explanation for BERT, there are dis-joint 500 samples used.

The aim of the evaluation is to observe the capability of the model explanation by PhilaeX in terms of the percentage of ``good'' explanations.
An adversarial sample has a ``good  explanation'',  only  if  a  certain number of the activated features in this sample are  attributed  with  positive  values.
A high number of ``activated features'' are identified in terms of their attribution values and means that the model explanation verifies the assumption that the model is evaded because of the activated features in the adversarial sample.

\begin{figure}
     \centering
     \begin{subfigure}[b]{0.5\textwidth}
         \centering
         \includegraphics[width=\textwidth]{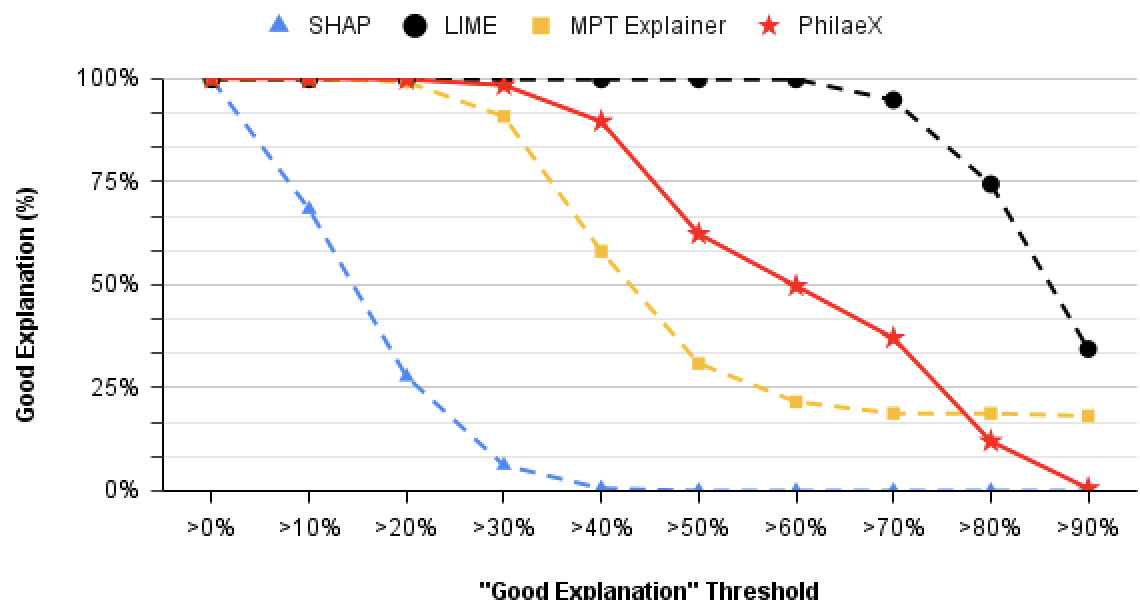}
         \caption{``Good explanation'' Percentage for SVM}
         \label{fig:y equals x}
     \end{subfigure}
     \hfill
     \begin{subfigure}[b]{0.5\textwidth}
         \centering
         \includegraphics[width=\textwidth]{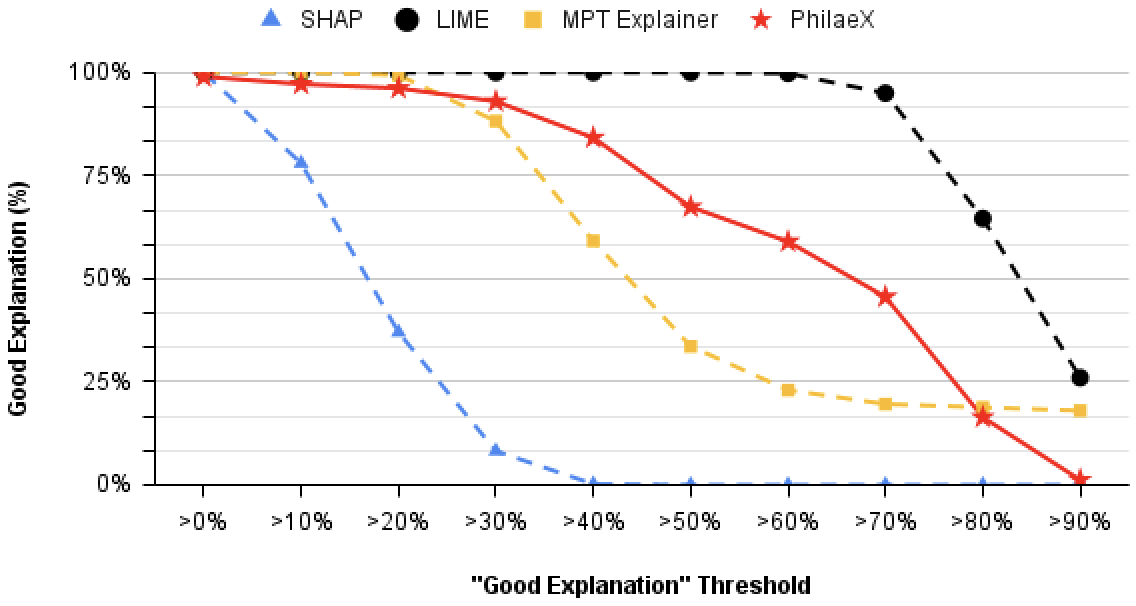}
         \caption{``Good explanation'' Percentage for BERT}
         \label{fig:three sin x}
     \end{subfigure}
        \caption{\textbf{``Good Explanation'' Comparison} The number of ``good explanations'' by PhilaeX stays in a high level (i.e., nearly 100\% in SVM), when the threshold of ``good explanation'' is less than 40\%. This also shows the robustness of the explanation by PhilaeX, compared to the other explainable AI methods.}
        \label{fig_adv_comp}
\end{figure}

In the experiment, we compare the explanation capability of PhilaeX against LIME~\cite{lime}, SHAP~\cite{NIPS2017_7062} and MPT explainer~\cite{mpt_explainer_2021}.
The reasons that we use these three explainable AI methods as the baseline are:
(1) LIME is a popular explainable AI method that explains the models by learning a linear surrogate model. 
(2) The explanation generated by SHAP is based on the computation of Shapley value~\cite{roth1988shapley}, which concept has been widely used in cooperative game theory.
(3) The recently MPT explainer is based on the modern portfolio theory~\cite{mpt} that was proposed in economics to allocate the investment to different assets for a maximum return with minimum risk.
In the evaluation, we vary the threshold of the ``good explanation'' from 0\% activated features in the adversarial samples identified to 90\% activated features identified.
This allows us to observe the robustness of the explanations from different explainable AI methods.
In Fig~\ref{fig_adv_comp}, it shows PhilaeX can identify more activated features from the adversarial samples compared to LIME, MPT explainer and SHAP, when the same threshold of ``good explanation'' is used and the threshold value is less than 40\% in SVM and 20\% in BERT.
In addition, PhilaeX's explanation shows much robustness that is verified by the slower decreasing curve, compared to SHAP and MPT explainer.
This conclusion still holds true when we compare the robustness of PhilaeX and LIME, considering the unstable explanation in LIME that is caused by the random perturbation on the features' values.

In the explanation of BERT, PhilaeX shows slight lower ratio of ``good explanations'', when the threshold of ``good explanation'' is less than 30\%.
This is possibly because BERT considers more joint contributions among the features that reduces the effect by single features accordingly.
However, we see that PhilaeX still presents a relatively robust explanation capability among these explainable AI methods, because of its slower curve decline.

\subsection{Explaining PDF Malware Detector}
\label{sec_sub_exp_pdf_malware_detection}
In the fidelity test, the aim is to evaluate if the explainer attributes high values for the features that has high impact on the model's prediction behavior.
Specifically, there are two kinds of tests we used in the experiments:
(1) \textbf{Deduction test} that removes a certain number of features with high attribution values will lead the model to predict the manipulated sample as the opposite class. 
That is, the less such high attribution value features are removed, the higher the explanation fidelity. 
For example, the SVM model predicts a manipulated sample of malware as non-malware, if the feature with the top attribution value is removed. This means that this feature is correctly attributed in the explanation;
(2) In \textbf{Augmentation test}, we activate a certain number of features in a non-malware sample. These features are from a malware sample and are attributed with high attribution values in the model explanation on this malware sample. It is expected that the model's prediction on the manipulated non-malware sample as malware, if the explanation is correct.
That is, the correctly attributed features in a malware sample may have strong individual impact on the model's prediction behavior that lead the model towards the malware class.

We use the positive classification rate (PCR)~\cite{guo2018lemna} as the evaluation metric to quantify the fidelity of the explanations.
The PCR is defined as the ratio of samples which retains their original class after the manipulation through deduction or augmentation.
The PCR in an explanation with high fidelity will be as low as possible through a deduction test, and as high as possible by the augmentation test.

\begin{figure*}
     \centering
     \begin{subfigure}[b]{0.45\textwidth}
         \centering
         \includegraphics[width=\textwidth]{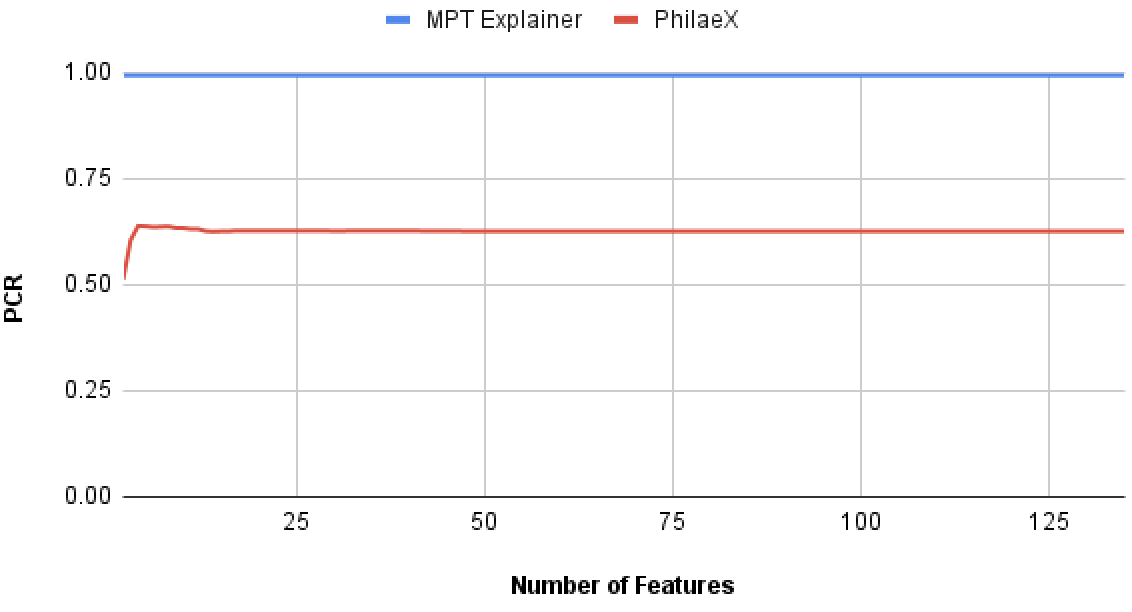}
         \caption{Deduction Test for Random Forest}
         \label{fig_ft_a}
     \end{subfigure}
     \hfill
      \begin{subfigure}[b]{0.45\textwidth}
         \centering
         \includegraphics[width=\textwidth]{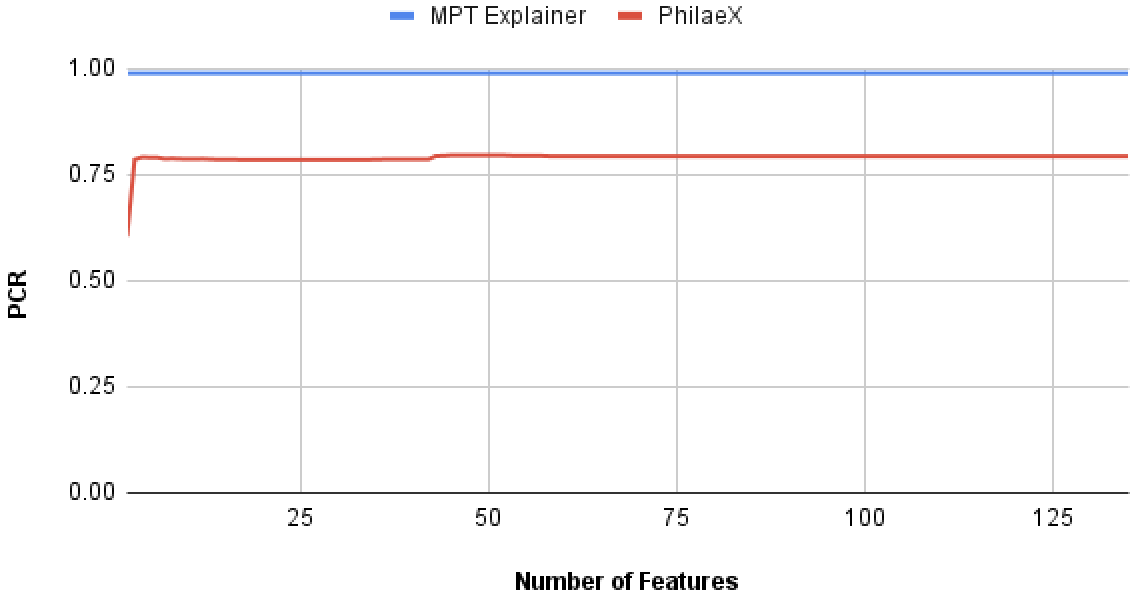}
         \caption{Deduction Test for SVM}
         \label{fig_ft_b}
     \end{subfigure}
     \hfill
          \begin{subfigure}[b]{0.45\textwidth}
         \centering
         \includegraphics[width=\textwidth]{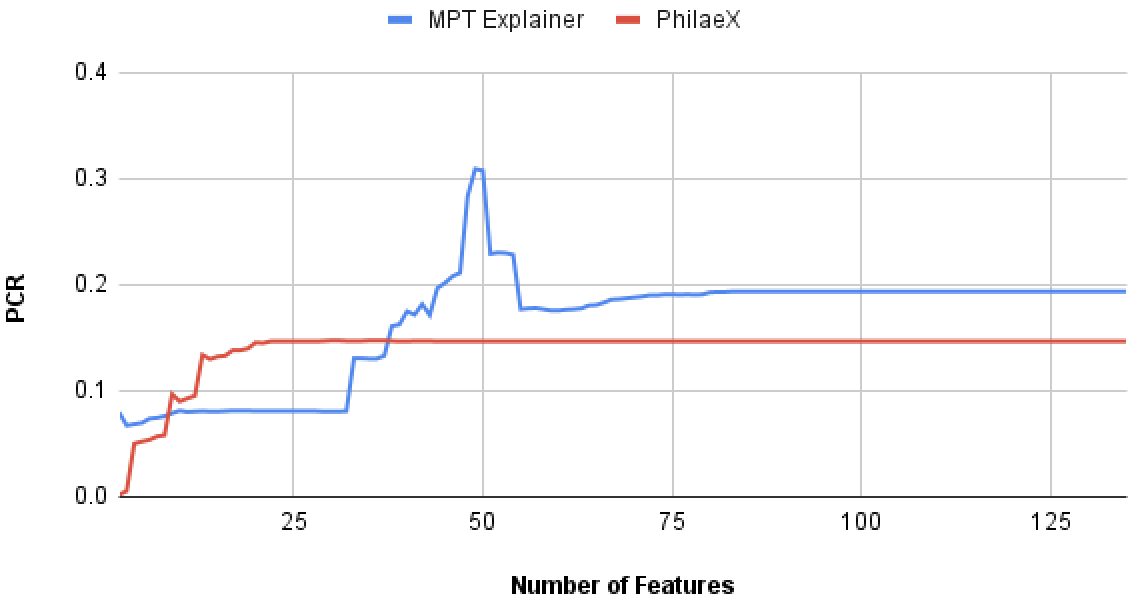}
         \caption{Augmentation Test for Random Forest}
         \label{fig_ft_c}
     \end{subfigure}
     \hfill
          \begin{subfigure}[b]{0.45\textwidth}
         \centering
         \includegraphics[width=\textwidth]{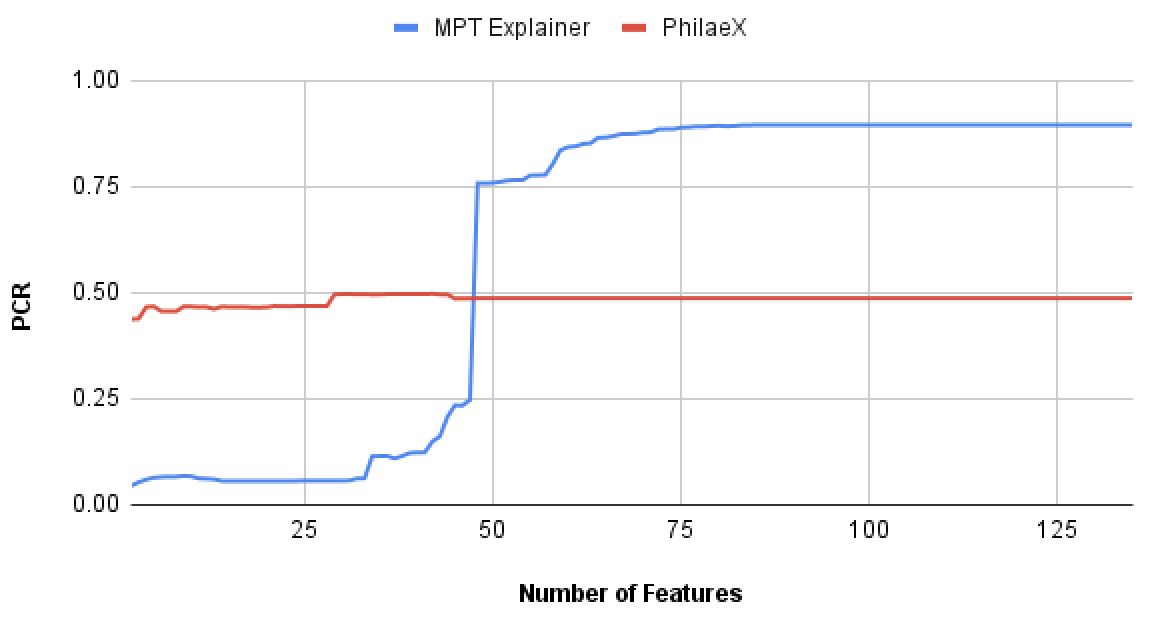}
         \caption{Augmentation Test for SVM}
         \label{fig_ft_d}
     \end{subfigure}
     
        \caption{\textbf{Fidelity Test} In the deduction test results (a) and (b), PhilaeX shows higher fidelity of explanation for both Random Forest and SVM classifiers, where the PCR value of deduction test should be as lower as possible. In the augmentation test, PhilaeX shows higher PCR values (better) for both Random Forest and SVM classifiers, when a small number of features used (i.e., $<$ 30 features in RF and $<$ 50 features in SVM). }
        \label{fig_fidelity_test}
\end{figure*}

In this experiment, we test the explanation fidelity by PhilaeX, when it is used to explain the Random Forest and SVM classifiers on the PDF malware detection task.
In Fig.~\ref{fig_ft_a} and Fig.~\ref{fig_ft_b}, we can observe that for both RF and SVM, PhilaeX has a significant higher fidelity explanation than that of MPT explainer, which are measured by the lower PCRs.
This finding verifies that the features selection function $h(\mathbf{x})$ in Section~\ref{sec_method_philaex} guarantees the following features attribution to assign high attribution values to the important features.
In addition, the high fidelity (in terms of PCR) is stable although the number of features used is increasing.
This means that PhilaeX is more capable of identifying the important features (by attributing it with higher value) than that of MPT explainer.

In Fig.~\ref{fig_ft_c} and Fig.~\ref{fig_ft_d}, the features with high attribution values by PhilaeX will generally guarantee a high PCR for both RF and SVM, when the number of features used are small.
However, the PCRs for PhilaeX are getting lower than that of MPT explainer when around 50 and more features are used in the augmentation test.
This is probably due to the joint contribution by all the features becoming stronger as the number of features used increases.

\subsection{Running Time Performance}
The average running time to explain the SVM's prediction on a single data sample of Android malware apps is around 6.37 seconds, compared to the MPT explainer with around 15.44 seconds.
This is probably due to the efficient the optimization process of Ridge regression.

\section{Conclusion}
\label{sec_conclusion}
In this article, we presented a novel model-agnostic explainable AI method, PhilaeX, that is featured by the features selection strategy and more suitable to explain the AI models used in cyber security tasks.
The explanation is in the form of features attribution for machine learning classifiers.
This method has a multi-stage feature selection function that identifies the candidate features to be explained:
(1) the core features to find the features that lead the model to make a borderline prediction;
(2) the features with positive individual contributions towards the model's prediction on the original sample to restrict the explainer to focus on important features' attribution, which is helpful in revealing the model's behavior in a more accurate way;
and (3) the Ridge regression model as the surrogate model quantifies the contributions of these features, considering the joint contributions made by them.
The explanation fidelity of the proposed method is evaluated by two experiments.
The first experiment aims to find the activated features from the adversarial sample of Android malware, through the attribution values (positive values) by PhilaeX.
The results shows PhilaeX has higher capability of the identification on such activated features than those by LIME, SHAP and MPT Explainer.
The second experiment consists of two fidelity tests, which are the deduction test and augmentation test.
In the deduction test, PhilaeX has significantly higher fidelity explanations than that of the MPT explainer.
The augmentation test reveals that PhilaeX has higher PCRs when a small number of features used.
Both experiments results show that PhilaeX can be helpful for explanation of the AI models, such as those used in the cyber security field.

\bibliographystyle{apalike}
{\small
\bibliography{philaex_arXiv_Zhi_Lu_20220702}}

\end{document}